\documentclass{camera}
\usepackage{graphicx}  
\begin{document}

%
\title{Advancement in the understanding of\\ multifragmentation
and phase transition\\ for hot nuclei}

%
\author{B. Borderie$^1$, E. Bonnet$^{1,2}$, F. Gulminelli$^3$, N. Le
Neindre$^{1,3}$,\\
D. Mercier$^{3,4}$, S. Piantelli$^5$, Ad. R. Raduta$^6$ \and M. F. Rivet$^1$ \\
for the INDRA and ALADIN collaborations}

%
\organization{
$^1$Institut de Physique Nucl\'eaire, CNRS/IN2P3, Univ. Paris-Sud 11,\\ Orsay,
France\\
$^2$GANIL, DSM-CEA/CNRS-IN2P3,Caen, France\\ 
$^3$LPC Caen, CNRS/IN2P3, ENSICAEN, Univ. de Caen, Caen, France\\	 
$^4$Institut de Physique Nucl\'eaire, IN2P3-CNRS et Universit\'e Lyon I,\\ Villeurbanne,
France\\
$^5$Sezione INFN, Sesto Fiorentino (Fi), Italy\\	 
$^6$National Institute for Physics and Nuclear Engineering,\\ Bucharest-M\u{a}gurele, Romania}

\maketitle

\begin{abstract}
Recent advancement on the knowledge of
multifragmentation and phase transition for hot nuclei
is reported. It concerns i) the
influence of radial collective energy on fragment partitions and the
derivation of general properties of partitions in presence of such a
collective energy, ii) a better
knowledge of freeze-out properties obtained by means of a simulation based
on all the available experimental information and iii) the quantitative study
of the bimodal behaviour of the heaviest fragment charge distribution for
fragmenting hot heavy quasi-projectiles which allows, for the
first time, to estimate the latent heat of the phase transition.
\end{abstract}

%
\section{Introduction}
Nucleus-nucleus collisions at intermediate energies offer various possibilities to
produce hot nuclei which undergo a break-up into smaller pieces, which is
called multifragmentation. The measured fragment properties are 
expected to reveal the existence on a phase transition for hot
nuclei which was earlier theoretically predicted for nuclear
matter~\cite{I46-Bor02,WCI06,Bor08}. 
By comparing in detail the properties of fragments ($Z\geq5$) emitted
by hot nuclei formed in central (quasi-fused systems, QF, from
$^{129}Xe$+$^{nat}Sn$, 25-50 AMeV)
and semi-peripheral collisions (quasi-projectiles, QP, from 
$^{197}Au$+$^{197}Au$, 80 and 100 AMeV),
i.e. with different dynamical conditions for their formation,
the role of radial collective
energy on partitions is emphasized~\cite{I69-Bon08} and general properties
of partitions are deduced (section 2).
Then, in section 3,
freeze-out properties of multifragmentation events produced in central
collisions ($^{129}Xe$+$^{nat}Sn$) are estimated~\cite{I66-Pia08} and confirm the existence of
a limiting excitation energy for fragments around 3.0-3.5 MeV per nucleon.
The deduced freeze-out volumes are used as a calibration to 
calculate freeze-out volumes for QP sources; thus one can locate where the different sources
break in the phase diagram.
Finally, in section 4, the charge distribution of the heaviest fragment
detected in the decay of QP sources is observed to be bimodal.
This feature is expected as a generic signal of phase transition in
nonextensive systems such as finite systems. For the first time an estimate
of the latent heat of the transition is also extracted~\cite{I72-Bon09}.
\section{Radial collective energy and fragment partitions}
To make a meaningful comparison of fragment properties which can be 
related to the phase diagram, hot nuclei showing, to a
certain extent, statistical emission features must be selected.
For central collisions (QF events) one selects complete and
compact events in velocity
space (constraint of flow angle $\geq 60^{\circ}$).
For peripheral collisions (QP subevents) the selection method
applied to quasi-projectiles minimizes the contribution
of dynamical emissions by imposing a compacity of fragments in velocity space.
The excitation energies of the different hot nuclei produced are calculated
using the calorimetry procedure (see~\cite{I69-Bon08} for details).
\begin{figure}[htb]
\begin{minipage}[c]{.45\textwidth}
\centering
\includegraphics[width=1.05\textwidth]
{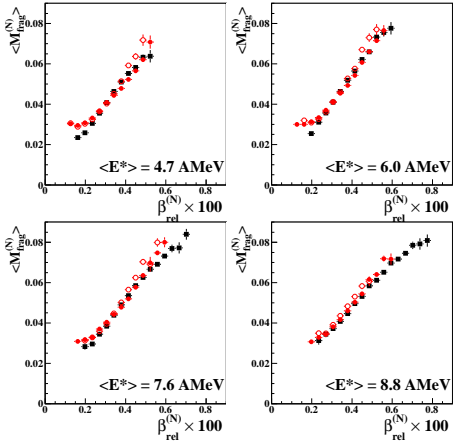}
\end{minipage}
\hspace{.05\textwidth}%
\begin{minipage}[c]{.45\textwidth}
\centering
\caption{
Evolution of the average fragment
multiplicity normalized to the source charge/size $<M^{(N)}_{frag}>$=
$<M_{frag}/Z_s>$  as a function of the relative velocity of fragments,
$\beta^{N}_{rel}$, (see text) for different total excitation energy per
nucleon of the sources.Full squares, open and full circles stand
respectively for QF sources and QP
sources produced at 80 and 100 MeV/nucleon incident energies.}
\label{fig1}
\end{minipage}
\end{figure}
By comparing the properties of selected sources on the same excitation
energy domain, significant differences are observed above 5 AMeV excitation
energy. QF sources have larger mean fragment
multiplicities, $<M_{frag}>$, even normalized to the sizes of the sources
(which differ by about
20\% for QF and QP sources), and lower values for generalized asymmetry:
$ A_{Z}= \sigma_{Z} / (\langle Z \rangle \sqrt{M_{frag}-1})$.
A possible explanation of those different fragment partitions
is related
to the different dynamical constraints applied to the hot nuclei produced:
a compression-expansion cycle for central collisions and a more gentle friction-abrasion
process for peripheral ones. 

\begin{figure}[htb]
\begin{minipage}[c]{.45\textwidth}
\centering
\includegraphics[width=1.05\textwidth]
{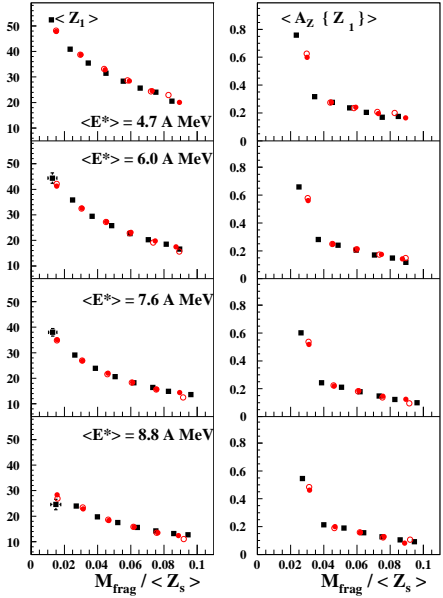}
\end{minipage}
\hspace{.05\textwidth}%
\begin{minipage}[c]{.45\textwidth}
\centering
\caption{
Left and right sides refer respectively to the mean charge of the
heaviest fragment of partitions, $<Z_1>$,
and to the generalized asymmetry in charge of the fragment partitions without
the heaviest one, $A_{Z} \backslash \{Z_{1}\}$, (see text) as a function of the reduced fragment
mutiplicity, $M_{frag}$/$<Z_s>$, for different total excitation energy per
nucleon of the sources.Full squares, open and full circles stand
respectively for QF sources and QP
sources produced at 80 and 100 MeV/nucleon incident energies.}
\label{fig2}
\end{minipage}
\end{figure}
The occurence of radial collective energy following a compression phase is 
predicted in semi-classical simulations of central collisions 
in the Fermi energy domain~\cite{Mol88,Surau89}. In experiments
it was recognized, in most of the cases, from
comparisons of kinetic properties of fragments with models.
The mean relative velocity between fragments, $\beta_{rel}$, 
independent of the reference 
frame, allows to compare radial collective energy
for both types of sources (QF or QP). 
The effect of the source size (Coulomb contribution on fragment velocities)
can be removed  by using a simple normalization 
which takes into account, event by event,
the Coulomb influence, in velocity space, of the mean fragment charge, 
$\langle Z \rangle$, on the complement of the source charge ($Z_{s}-\langle Z \rangle$):
$ \beta^{(N)}_{rel} = \beta_{rel} / \sqrt{\langle Z \rangle (Z_{s}- 
\langle Z \rangle)}$. At
an excitation energy of about 5 AMeV, the $\beta^{(N)}_{rel}$ values
corresponding to QF and QP sources are similar.
Above that excitation energy, values for QF sources exhibit a strong
linear increase, whreras for QP sources $\beta^{(N)}_{rel}$ only slightly increases
up to 9-10 AMeV. 
That fast divergence between the values of
$\beta^{(N)}_{rel}$ for the two types of sources signals the well known 
onset of radial collective expansion for central collisions. 
In~\cite{I40-Tab03}, estimates of radial collective energy 
(from 0.5 to 2.2 AMeV) for QF sources produced by Xe+Sn collisions are 
reported for four incident energies: 32, 39, 45 and 50 AMeV.
Those estimates which were extracted from comparisons with the statistical model
SMM assuming a self similar expansion energy have
been used to calibrate the $\beta^{(N)}_{rel}$ observable
(see~\cite{I69-Bon08} for details).
Then, one can show that radial collective energy is 
essentially due to thermal pressure for QP sources
in semi-peripheral heavy-ion collisions as it is in hadron-induced
reactions~\cite{Beau01}. 
For QF sources produced in central heavy-ion collisions the contribution 
from the compression-expansion cycle becomes more and more important 
as the incident energy increases.  
Figure~\ref{fig1} shows, for different total excitation energy per nucleon
corresponding to values
defined by QF sources, the evolution of the average fragment
multiplicity normalized to the source charge/size $<M^{(N)}_{frag}>$=
$<M_{frag}/Z_s>$  as a function of $\beta^{N}_{rel}$. We observe a well
defined correlation which fully confirms the role of collective energy in
producing more fragments. Depending on the source type the
relative contributions to radial collective energy of thermal pressure
and cycle compression-expansion strongly differ~\cite{I69-Bon08}
but anyhow $\beta^{N}_{rel}$,
representative of the total collective energy, fixes the average degree
of fragmentation (normalized mean fragment multiplicities).

Does the intensity of the radial collective energy also
govern the details of fragment partitions, namely
the relative charge/size of fragments in partitions.
One can first consider the evolution of the
size of the heaviest fragment, for given total excitation energies, with
the reduced fragment multiplicities  $M_{frag}$/$<Z_s>$.
On the left panel of fig.~\ref{fig2}
average values of the heaviest fragment charge for QP and QF
sources are reported: they  follow exactly the same evolution. Finally the
division of
the charge among other fragments is investigated using the generalized asymmetry in charge
of the fragment partitions.
One can re-calculate the generalized asymmetry by removing
$Z_{1}$ from partitions, noted 
$A_{Z} \backslash \{Z_{1}\}$. The results, displayed in fig.~\ref{fig2} (right
panel) do not depend on the source type at a given
total excitation energy and a given reduced fragment multiplicity.
Note that the general asymmetry follows a linear trend except the lower
reduced fragment multiplicity which corresponds to $M_{frag}$=2; indeed in that
case, after removing $Z_1$, only one fragment is available for the asymmetry
calculation and in each event the heaviest Z of the partition below Z equal 5
was taken.
Such a result shows the subtle role played by the radial collective
energy. It influences the overall degree of fragmentation but it does not affect the
relative size of
fragments in partitions for fixed reduced fragment multiplicities. Those
results represent a benchmark against which models describing fragmentation
of finite systems should be tested.
\section{Freeze-out properties}
\begin{figure}[htb]
\begin{center}
\includegraphics*[width=0.9\textwidth]
{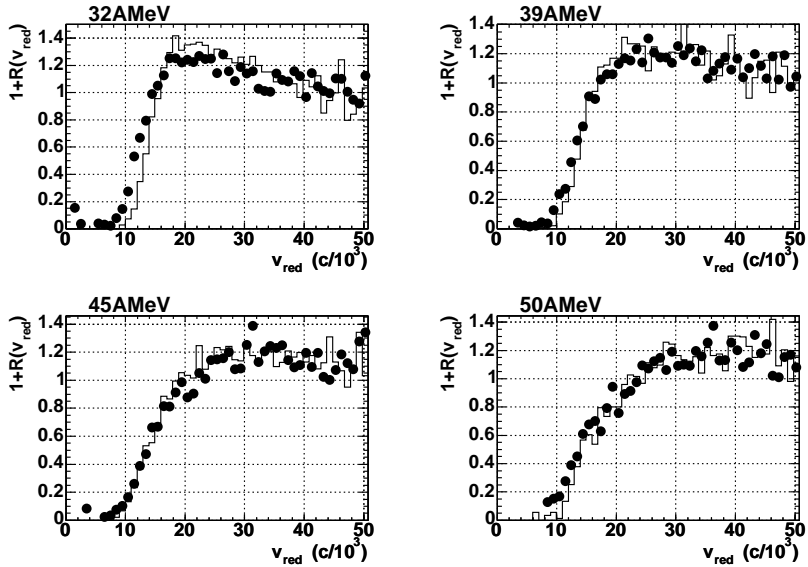}
\caption{Comparison between the experimental (full points) and simulated
(histograms) reduced relative velocity correlation functions for
all the fragments. The
reduced relative velocity between two fragments with charges $Z_i$ and $Z_j$
($Z_{i,j}>$4) is
defined as $v_{red}$=$v_{rel}$/$(Z_i+Z_j)^{1/2}$. Each panel 
refers to a different beam energy: 32 AMeV (top left),
39 AMeV (top right), 45 AMeV (bottom left) and 50 AMeV
(bottom right). From~\protect\cite{I66-Pia08}.}
\label{fig3}
\end{center}
\end{figure}

Starting from all the available experimental information of 
selected QF
sources produced in central $^{129}$Xe+$^{nat}$Sn collisions which undergo
multifragmentation, a simulation was performed to reconstruct freeze-out
properties event by event~\cite{I58-Pia05,I66-Pia08}. 
The method requires data with a very high degree
of completeness, which is crucial for a good estimate of Coulomb energy.
The parameters of the simulation were fixed in a
consistent way including experimental partitions, kinetic properties and
the related calorimetry. The necessity of introducing a limiting temperature
related to the vanishing of level density for fragments~\cite{Koo87}
in the simulation was confirmed for all
incident energies. This naturally leads to a limitation of their excitation
energy around 3.0-3.5 AMeV as observed in~\cite{I39-Hud03}.
The agreement between experimental and simulated velocity spectra for
fragments of given charges (Z=6, 11, 18 and 27), for the
different beam energies, is quite remarkable 
(see~\cite{I66-Pia08}). Finally
relative velocities between fragment pairs were also compared
through reduced relative velocity correlation 
functions~\cite{Kim92,Bow95,Gro97,I57-Tab05}
(see fig.~\ref{fig3}).
In the simulation the fragment
emission time is by definition equal to zero and correlation
functions are consequently only sensitive 
to the spatial arrangement of fragments at break-up and the
radial collective energy involved
(hole at low reduced
relative velocity), to source sizes/charges and to
excitation energy of the sources (more or less pronounced bump
at $v_{red}$= 0.02-0.03c).
Again a reasonable agreement is obtained between experimental data and
simulations, especially at 39 and 45 AMeV incident energies, which
indicates that the retained method and parameters are sufficiently relevant
to correctly describe freeze-out topologies and properties.

 The major properties of the freeze-out configurations
thus derived are the following: an important increase, 
from $\sim$20\% to $\sim$60\%, of
the percentage of particles present at freeze-out between 32 and 45-50 AMeV 
incident energies accompanied by a weak increase of the freeze-out volume 
which  tends to saturate at high excitation energy.  
Finally, to check the overall physical coherence of the developed approach,
a detailed comparison with a microcanonical statistical model (MMM) was
done. The degree of agreement, which was found acceptable, confirms the main
results and gives confidence in using those reconstructed freeze-out events
 for further studies as it is done in~\cite{I69-Bon08}.
Estimates of freeze-out volumes for QF sources produced 
in Xe+Sn collisions for incident energies between 32 and 50 AMeV
evolve from 3.9 to 5.7 $V/V_0$, where $V_0$ would correspond to 
the volume of the source at normal density~\cite{I66-Pia08}. 

\begin{figure}[htb]
\begin{minipage}[c]{.45\textwidth}
\centering
\includegraphics[width=1.15\textwidth]
{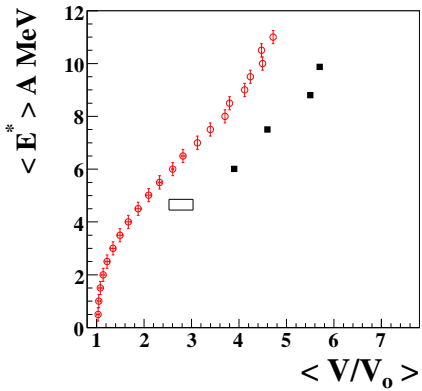}
\end{minipage}
\hspace{.05\textwidth}%
\begin{minipage}[c]{.45\textwidth}
\centering
\caption{Fragmentation position in the excitation energy-freeze-out volume
plane. The four full squares (QF sources) are taken from~\protect\cite{I66-Pia08}. 
The open rectangle gives the estimated position (with error bar) for QF
source at 25~AMeV, and the open circles those for QP sources.
From~\protect\cite{I69-Bon08}.}\label{fig4}
\end{minipage}
\end{figure}

To calibrate the freeze-out volumes for other sources,
we use the charge of the heaviest fragment $<Z_1^{(N)}>$ or the 
fragment multiplicity $<M_{frag}^{(N)}>$, normalized to the size of the 
source, as representative of the volume or density at break-up.
From the four points for QF sources and the additional constraint that
$Z_1^{(N)} = M_{frag}$=1 at $V/V_0$=1, we obtain two relations
$V/V_0 = f_1(Z_1^{(N)} )$ and  $V/V_0 = f_2(M_{frag}^{(N)})$, from which we
calculate the volumes for QF sources at 25~AMeV and for QP sources.
The results are plotted in fig.~\ref{fig4}, with error bars coming
from the difference between the two estimates using $f_1$ and $f_2$; note that 
error bars for the QP volumes are small up to 7~AMeV, and can not be 
estimated above, due to the fall of $<M_{frag}^{(N)}>$ at high energy (see
fig. 5 of~\cite{I69-Bon08}). So only $<(Z_1^{(N)}>)$ can be used
over the whole excitation energy range considered and the derived function
is the following:\\$V/V_0 = exp(2.47-4.47<(Z_1^{(N)}>)+0.86$. 

The volumes of QP sources are smaller than those of QF sources
(by about 20\% on the $E^*$ range 5-10~AMeV). 

$Z_1$ also presents some specific dynamical properties. As shown 
in~\cite{I9-Mar97,I57-Tab05} for QF sources, 
its average kinetic energy is smaller than that of other fragments with the
same charge. The effect was observed whatever the fragment multiplicity for
Xe+Sn between 32 and 50 AMeV and for Gd+U at 36 AMeV. The
fragment-fragment correlation functions are also different when one
of the two fragments is $Z_1$.
This observation was
connected to the event topology at freeze-out, the heavier 
fragments being systematically closer to the centre of mass than the 
others.

\section{Bimodality of the heaviest fragment and latent heat of
the transition}
At a first-order phase transition, the distribution of the order parameter 
in a finite system presents a characteristic bimodal behaviour in the 
canonical or grandcanonical ensemble~\cite{Gul03}.
The bimodality comes from an anomalous convexity of the underlying 
microcanonical entropy~\cite{Gros02}. It physically corresponds to the 
simultaneous presence of two different classes of physical states for the same 
value of the control parameter, and can survive at the thermodynamic 
limit in a large class of physical systems subject to long-range 
interactions~\cite{LNP02}.
\begin{figure}[!hbt]
\begin{minipage}[c]{0.45\textwidth}
\centering
\includegraphics[trim = 0 0 0 1,clip,width=\textwidth]
{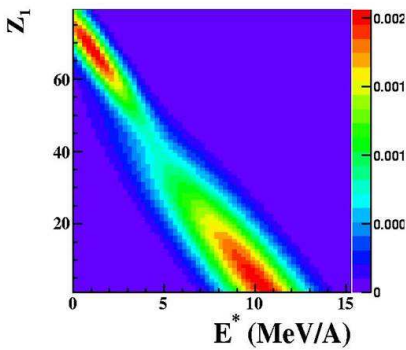}
\end{minipage}%
\hspace{.05\textwidth}%
\begin{minipage}[c]{0.45\textwidth}
\centering
\caption{
Size of the heaviest fragment versus total excitation energy in AMeV. That picture
is constructed using the fit parameters extracted from the
equivalent-canonical distribution. The distance
between the two maxima, liquid and gas peaks, projected on the excitation energy
axis corresponds to the latent heat of the transition.}
\label{fig5}
\end{minipage}
\end{figure}
In the case of hot nuclei which undergo multifragmentation, the size/charge
of the heaviest fragment was early recognized as an order
parameter~\cite{Bot01,I51-Fra05} using the universal fluctuation theory.
In a quantitative analysis for QP sources, 
the robustness of the signal of bimodality is tested 
against two different QP selection methods~\cite{I72-Bon09}. A weighting 
procedure~\cite{Gul07} is used to test the independence of the 
decay from the dynamics of the entrance channel and to allow a comparison
with canonical expectations. Finally, a double 
saddle-point approximation is applied to extract from the measured data 
an equivalent-canonical distribution. 
To take into account the small variations
of the source size, the charge of the heaviest fragment $Z_1$ has been normalized
to the source size. After the weighting procedure, a bimodal behaviour of the largest
fragment charge distribution is observed for both selection methods.
Those weighted experimental distributions can be fitted with an
analytic function (see~\cite{I72-Bon09} for more details).
From the obtained parameter values one can estimate the
latent heat of the transition of
the hot heavy nuclei studied (Z$\sim$70) as 
$\Delta E=8.1 (\pm0.4)_{stat} (+1.2 -0.9)_{syst}$~AMeV.
Statistical error was
derived from experimental statistics and systematic errors
from the comparison between the different QP selections.
The results (for one QP source selection) are illustrated in
fig.~\ref{fig5}. A detailed presentation and discussion  of those results
is also found in~\cite{BonIWM}.
\section{Conclusion}
Today a rather coherent and complete picture has been reached for a few
exhaustive studies concerning multifragmentation and the related liquid-gas
type phase transition which occurs at excitation energies between
2-3 and 9-10 AMeV. Only the mechanism of fragment formation is still an
open question: spinodal fluctuations in stochastic mean field approaches or
many-body correlations early built in molecular dynamics~\cite{Bor08}.
At present, with the introduction of the N/Z degree of freedom in such
studies, new signals predicted by theory must be investigated to precise and
strengthen our actual knowledge. It concerns fractionation (liquid more
symmetric due to minimization of symmetry energy in dense phase), increased
fractionation if spinodal instabilities are responsible for fragment
formation, and the reduction of spinodal zone for large N/Z values involved. 
For the future the introduction of the N/Z degree of freedom will permit to
improve fundamental information on the phase diagram of baryonic matter.
Moreover hot exotic nuclei appear as a unique laboratory to serve as
test-bench for theory of phase transitions of finite quantum systems with
two components.


%
\end{document}